\documentclass[runningheads]{llncs}
\usepackage[T1]{fontenc}
\usepackage{graphicx}
\usepackage[breaklinks]{hyperref}
\usepackage{xurl}

\usepackage[acronym]{glossaries}
\usepackage[htt]{hyphenat}
\usepackage{listings}
\usepackage[style=numeric]{biblatex}
\usepackage{booktabs}
\usepackage{subfiles} 

\newacronym[longplural=Knowledge Graphs]{kg}{KG}{Knowledge Graph}
\newacronym{dkg}{DKG}{Decentralized Knowledge Graph}
\newacronym{sssom}{SSSOM}{Simple Standard for Sharing Ontological Mappings}
\newacronym{iri}{IRI}{Internationalized Resource Identifier}
\newacronym{ltqp}{LTQP}{Link Traversal Query Processing}
\newacronym{rdf}{RDF}{Resource Description Framework}
\newacronym{http}{HTTP}{Hypertext Transfer Protocol}
\newacronym{https}{HTTPS}{Hypertext Transfer Protocol Secure}

\glsdisablehyper

\usepackage[dvipsnames,svgnames]{xcolor} 
\usepackage[normalem]{ulem} 

\makeatletter
\font\uwavefont=lasyb10 scaled 700
\def\spelling{\bgroup\markoverwith{\lower3.5\p@\hbox{\uwavefont\textcolor{Red}{\char58}}}\ULon}
\def\grammar{\bgroup\markoverwith{\lower3.5\p@\hbox{\uwavefont\textcolor{LimeGreen}{\char58}}}\ULon}
\def\phrasing{\bgroup\markoverwith{\lower3.5\p@\hbox{\uwavefont\textcolor{RoyalBlue}{\char58}}}\ULon}

\newcommand\remove{\bgroup\markoverwith{\textcolor{red}{\rule[0.5ex]{2pt}{0.4pt}}}\ULon}
\newcommand\insertion{\bgroup\markoverwith{\textcolor{Green}{\rule[-0.5ex]{2pt}{0.6pt}}}\ULon}
\makeatother

\begin{document}

\title{Demonstrating Online Schema Alignment in Decentralized Knowledge Graphs Querying}
\titlerunning{Demonstrating Online Schema Alignment}
%
%
\author{Bryan-Elliott Tam\inst{1}\orcidID{0000-0003-3467-9755} \and
Pieter Colpaert\inst{1}\orcidID{0000-0001-6917-2167} \and
Ruben Taelman\inst{1}\orcidID{0000-0001-5118-256X}}
\authorrunning{B-E. Tam et al.}
%
\institute{Ghent University, Gent Belgium
}

\institute{
  IDLab,
  Dept.\ Electronics and Information Systems\,
  Ghent University -- imec, Belgium
}

\maketitle

\begin{abstract}
Decentralized \acrlongpl{kg} querying enables integrating distributed data without centralization, but is highly sensitive to vocabulary heterogeneity.
Query issuers cannot realistically anticipate all vocabulary mismatches, especially when alignment rules are local, scoped, or discovered at runtime.
We present an \emph{online schema alignment} approach for \acrfull{ltqp} that discovers, scopes, and applies alignment rules dynamically during query execution while preserving traversal behavior.
This demo paper demonstrates the approach on a decentralized social-media scenario through a web interface built on a Comunica-based \acrshort{ltqp} engine.
Source code, a CLI, and a reusable library are publicly available.
The demonstration shows that online schema alignment recovers complete query results with low overhead, providing a practical foundation for web-scale reasoning in \acrshort{ltqp} systems.
\end{abstract}


\section{Introduction}

Integrating multiple \glspl{kg}~\cite{linked_data} presents scalability and heterogeneity challenges.
Data volume, decentralization across independently managed sources, or privacy requirements often necessitate federated or link traversal-based querying~\cite{Hartig2012} rather than centralized materialization.
\Gls{ltqp}~\cite{Hartig2012} is particularly suited to such settings as it exploits the fact that IRIs in triples can be dereferenced to discover further \glspl{kg}.
However, in unindexed distributed networks, different providers often express semantically equivalent information using heterogeneous vocabularies, and query issuers expect to formulate queries under a single, consistent representation.
Schema alignment, which maps knowledge from one vocabulary to a semantically equivalent representation in another, can simplify the formulation of such queries.
Prior work has investigated reasoning over SPARQL endpoints~\cite{terdjimi2016hylar}, RDF streams~\cite{bonte2025languages}, and schema alignment in SPARQL systems~\cite{cheng2023considering, Joshi2012, maarten2023pod}, but existing approaches assume alignment rules are known and scoped a priori.
This assumption is unrealistic in large decentralized networks, given the complexity of ontology design and the possibility of provider-side modeling mistakes~\cite{abedjan2012reconciling, lorey2011rdf}.
Even trivial inconsistencies (e.g., HTTP vs.\ HTTPS schema.org vocabularies) require query rewriting such as \texttt{UNION} clauses.
Deeper semantic mismatches, for instance where providers conceptualize responses as ``replies'' or ``comments'', or represent attributes as literals or IRIs, are even harder.
Rule chaining compounds the problem by requiring either materializing intermediate \glspl{kg} through multiple \texttt{CONSTRUCT} queries or backchaining to produce a final query~\cite{maarten2023pod}.
Both methods are incompatible with \gls{ltqp}, as the former loses awareness of previously traversed documents and the latter fails to discover intermediate resources needed for traversal.
Our work introduces \textit{online schema alignment}, where schema mappings are discovered, scoped, and applied dynamically during \gls{ltqp} query execution.
The rest of this paper presents \hyperref[sec:method]{our approach}, \hyperref[sec:demo]{the demonstrator}, and \hyperref[sec:conclusion]{conclusions}.

\section{Online Schema Alignment}\label{sec:method}

Schema alignment requires the definition of a set of rules.
In this work, we draw\footnote{-- Vocabulary: \url{https://onlineschemaalignmentltqp.github.io/vocabulary/vocab.ttl} -- Example: \url{https://onlineschemaalignmentltqp.github.io/vocabulary/example.ttl}}
on the \gls{sssom} specification~\cite{sssom_website, matentzoglu2022simple} to structure and formalize these rules. 
For the purpose of this work, we developed a simplified version of \gls{sssom} and use the concept of \emph{subweb}~\cite{traveling_map_ltqp}, a sub-\gls{dkg} defined by the \gls{kg} derived from a set of \glspl{iri} controlled by a data provider, to scope the domain of applicability of the alignment rules.
We opted for this design for several practical reasons.
First, our system operates under the open-world assumption, in which multiple new \glspl{kg} and schema alignment rules may be discovered during processing.
To ensure robustness, the system should prevent infinite recursive rule applications.\footnote{With our schema alignment entailment, logical contradictions are not possible.}
When such situations occur, it should ideally resolve them without aborting the entire computation.
Since rules are applied within specific domains, the likelihood of infinite recursion is significantly reduced.
In cases where infinite recursion is detected or overlap of subwebs, the simple resolution strategy of rejecting the newly generated schema alignment rule is applied.
Second, there is the potential misuse, overspecification, and underspecification of ontology terms~\cite{abedjan2012reconciling, lorey2011rdf}.
In our use case, the goal is to enable querying data across the web under the assumption that data providers best understand their own data models and the intended entailments.
Thus, we scope the rules to non-overlapping subwebs to prevent unintended entailments.
However, we also allow query issuers to define additional rules that may interact with those provided by data publishers.
This is justified by the fact that query issuers can more easily track and manage the consequences of their own entailment definitions, whereas it would be unrealistic to expect data providers to anticipate the broader implications of their rules beyond the scope of their own datasets or use cases.
Third, global alignment rules that are discoverable and executable on the web present a potential attack vector for malicious actors.

In \gls{ltqp}, a traversal policy determines which discovered \glspl{iri} the engine should dereference to find additional \glspl{kg}.
To discover alignment rules, we extend this policy to also follow object terms of triples with predicate IRI \texttt{semmap:\\ruleSetLocation}, where the dereferenced term provides a rule set.
When the engine encounters a rule set, it applies the rules to its internal \gls{kg} via forward chaining, producing additional aligned triples.
These aligned triples, rather than the original ones, are passed to the traversal policy and used for join processing, as traversal policies are not designed to handle vocabulary mismatches.

\section{Demonstration}\label{sec:demo}

We implemented an online schema alignment system using the link traversal version of the Comunica~\cite{taelman2018comunica} query engine.
Both our demonstration\footnote{\url{https://github.com/onlineSchemaAlignmentLTQP/demo}} and implementation\footnote{\url{https://www.npmjs.com/package/query-sparql-link-traversal-solid-schema-alignment}} are open source and publicly available.
A video of the demonstration is available online.\footnote{\url{https://www.youtube.com/watch?v=fGDQwu65los}}
Users can download the implementation to integrate it as a library or use the command-line interface to execute queries.
Our demonstration scenario is based on a social media network built using the SolidBench benchmark~\cite{Taelman2023}.
This social media network is decentralized, meaning that users store their data in personal data vaults called \emph{pods}; there is no centralized endpoint to query user information.
For the purpose of this demonstration, we modified the user data such that different users express similar information using different vocabularies.
Additionally, each user exposes a file that describes their schema alignment rules.
The demonstrator is a web-based application that allows users to execute either the proposed queries or arbitrary custom queries.
Queries can be run over two network configurations: the base network, where all pods use the same vocabulary, and the modified network, where pods use alternative vocabularies.
The proposed queries are typical social media-related queries, such as retrieving information about a user, identifying the forums where a user has posted, and finding posts liked by specific users.
To illustrate, consider a query retrieving a user's profile, when executed over the modified network without alignment, the query returns incomplete results because some pods use HTTPS variants or alternative terms for the same properties.
With online schema alignment enabled, the engine discovers the relevant mapping rules during traversal and returns the complete result set.
For each execution, the system provides detailed feedback, including the query results, execution time, and the alignment rules discovered during query processing, along with their associated subwebs.
Table~\ref{tab:eval} reports the execution times\footnote{\url{https://github.com/constraintAutomaton/Online-Schema-Alignment-for-Link-Traversal-Queries-in-Decentralized-Knowledge-Graphs/blob/main/eval/nomal_query_eval_times.md}} over ten runs per query.
For queries that complete under both configurations, alignment introduces an average overhead of $0.28 \pm 0.31$\,s.

\begin{table}[t]
  \centering
  \begin{tabular}{lrr}
    \toprule
    Query & With alignment (s) & Without alignment (s) \\
    \midrule
    Messages of liked users       & $6.90 \pm 1.58$ & --- \\
    Forums a user posted          & ---             & --- \\
    User information              & $1.09 \pm 0.07$ & $1.05 \pm 0.05$ \\
    Posts of a user               & $0.35 \pm 0.05$ & $*$ \\
    Tag distribution              & $2.77 \pm 0.22$ & $2.11 \pm 0.27$ \\
    \bottomrule
  \end{tabular}
  \caption{The usage of online schema alignment does not have a large impact on the query execution time of the demo queries. "---" indicates a timeout (180\,s) and "$*$" indicates incomplete results.}
  \label{tab:eval}
\end{table}
Users can alternate between configurations to compare result completeness and observe the overhead introduced by schema alignment.
Users can also define custom schema alignment rules scoped to specific subwebs, which can additionally serve to align the vocabulary of their queries to that of a target network.

\section{Conclusion}\label{sec:conclusion}
In this demo paper, we present a system for online schema alignment.
Through this demonstration, we show that, in unindexed and decentralized networks, \gls{ltqp} queries can be executed successfully in a generalized manner without increasing the complexity of the user-issued queries.
Limitations remain in the expressivity of those rules.
Future work includes extending the system to support more advanced reasoning mechanisms, such as the SPARQL entailment regimes, full \gls{sssom}, Notation3 and the upcoming SHACL rules.
Engines such as the EYE reasoner~\cite{n3Verborgh} or modern Prolog~\cite{Colmerauer1996} implementations like Scryer Prolog\footnote{\url{https://www.scryer.pl/}} could enable such reasoning capabilities.
More broadly, this work contributes to the vision of querying the web as a single coherent knowledge base, where data sovereignty and vocabulary diversity coexist with seamless interoperability.


\section{Acknowledgement}

This research was supported by SolidLab Vlaanderen (Flemish Government, EWI RRF project VV023/10) and Serendipity Engine (Research Foundation - Flanders (FWO) grant number S006323N).
Ruben Taelman is a postdoctoral fellow of the Research Foundation – Flanders (FWO) (1202124N).

\printbibliography

\end{document}